\def   \ni {\noindent}

\def   \ssk {\vskip  5truept}

\def   \bsk {\vskip 15truept}

\def   \newline {\hfil\break}

\documentstyle[epsfig]{article}
\begin{document}

\hsize 5truein
\vsize 8truein
\font\abstract=cmr8
\font\keywords=cmr8
\font\caption=cmr8
\font\references=cmr8
\font\text=cmr10
\font\affiliation=cmssi10
\font\author=cmss10
\font\mc=cmss8
\font\title=cmssbx10 scaled\magstep2
\font\alcit=cmti7 scaled\magstephalf
\font\alcin=cmr6 
\font\ita=cmti8
\font\mma=cmr8
\def\ref{\par\noindent\hangindent 15pt}
\null
%\vskip 3.0truecm
%\baselineskip = 12pt

% ------ beginning of font "title" ------

\title{\ni GAMMA-RAY BURST LOCALIZATION WITH \\    
INTEGRAL IN THE INTERPLANETARY NETWORK}                                               

% beginning of font "author and affiliation"
\bsk \bsk
\author{\ni K. Hurley $^{1}$}                                                       
\bsk
\affiliation{1) UC Berkeley, Space Sciences Laboratory, Berkeley, CA, 
94720-7450, USA}                                                
\bsk
\baselineskip = 12pt

% beginning of font "abstract and keywords"
\abstract{ABSTRACT \ni
We discuss the utilization of INTEGRAL as a gamma-ray burst detector in
the interplanetary network.  IBIS will have an inherent GRB detection
and localization capability.  However, the SPI anticoincidence shield will also be a
sensitive burst detector which, in conjunction with other spacecraft,
will make it possible to localize three to four times
more bursts than IBIS alone.  This capability will be quite unique in the years
2001-2003 when INTEGRAL is operating.
 }                                                 
\bsk
\baselineskip = 12pt
\keywords{\ni KEYWORDS: gamma-rays: bursts.
}               

\bsk
\baselineskip = 12pt

% beginning of font "text"

\text{\ni 1. INTRODUCTION
\ssk
\ni     

The events of the past year leave no doubt that the most promising avenue of research into the origins of cosmic gamma-ray bursts is the study of their multiwavelength counterparts.  This will require many $\approx$ arcminute-size error boxes.  Today, missions such as BeppoSAX, RXTE, and the Interplanetary Network can produce them.  HETE-II can do the same in the years 1999 - 2001, and a future GRB MIDEX mission, if selected, can revolutionize the study of GRB's in the years beyond 2003.  However, as table 1 demonstrates, there are currently no missions planned for the years 2001 - 2003, other than INTEGRAL, which have this capability.

\begin{table}[h]
\begin{center}
\caption{Table 1. Current and Future GRB Missions}
\end{center}
\vspace{0.4 in}
\begin{tabular}{lcccc}
MISSION & BURSTS/YR & ACCURACY & DELAY & DATES \\
BeppoSAX & 10 & 1-10' & Hours & Through 2000? \\
BATSE (GCN) & 300 & 6 - 20$^{\rm o}$ dia. & 5 s & Through 2002? \\
BATSE (Locburst) & 120 & 4 - 8$^{\rm o}$ dia. & 15-30 min & Through 2002? \\
Current IPN & 70 & 5'x10$^{\rm o}$ - 1'x20' & $\sim$day & Through 2001 at least \\
RXTE & 4 & 5' & hours & Through 2002? \\
HETE II & 24 & 10" - 10' & $\sim$5 s & 1999 - 2001 \\
INTEGRAL & 20 & $\sim$arcminutes & $\sim$5 - 100 s & 2001 - 2003 \\
MIDEX & $\>$100 & $\sim$arcseconds & $\sim$5 s & 2003 - 2005 \\
\end{tabular}
\end{table}
%       Text of 2nd paragraph

\bsk
\ni
2.1 INTEGRAL in the Interplanetary Network
\ssk
\ni

The inherent GRB capability of INTEGRAL in table 1 derives from the IBIS telescope (e.g. Pedersen et al. 1997).  However, this can be considerably enhanced by the addition of the INTEGRAL spacecraft to the Interplanetary Network.  

In an IPN, the interplanetary GRB experiments are constrained to have very low mass ($\approx$ 1 kg) and telemetry rates ($\approx$ 20 b/s).  This dictates the use of small detectors (typical areas $\approx$ 20 cm$^2$), so these instruments trigger only on the brighter bursts.  Near-Earth spacecraft are usually not limited in this way, so their instruments are considerably larger and more sensitive.  In the case of the 3rd IPN for example, each BATSE detector is $\approx$ 100 times larger than the Ulysses detector, and BATSE detects about one GRB/day.  By searching the untriggered Ulysses data for BATSE bursts, it is possible to effectively lower the detection threshold for Ulysses bursts, and $\sim$70 Ulysses/BATSE GRBs/y are detected this way (Hurley et al. 1998).  This is a key feature of any IPN: there should be a large, near-Earth detector with considerably greater sensitivity than can be achieved on the interplanetary spacecraft.

Several interplanetary missions are planned and proposed for the years 2001 - 2003.  These include:\\

1) \underline{Mars Surveyor Orbiter (figure 1)}  To be launched in 2001, NASA's MSO has two GRB detectors.  One is associated with the Ge spectrometer (W. Boynton, P.I.), and the other with the neutron detector (I. Mitrofanov, P.I.).  After a one year cruise, MSO will enter Martian orbit and return data for at least one Martian year (687 d)\\

2) \underline{Near Earth Asteroid Prospector (figure 2)} NEAP is a mission by the SpaceDev Corporation (a private company) to the asteroid Nereus, to be launched in 2001.  A small GRB detector has been proposed as a MIDEX mission of opportunity.  If accepted, NEAP would operate through the middle of 2003 and possibly beyond.\\

3) \underline{Ulysses} Now in its second orbit around the Sun, it has a nominal end of mission in December 2001.  However, a 3rd orbit will be proposed, and if accepted, the spacecraft could continue operations through 2006.  Ulysses carries a GRB detector which is part of the 3rd IPN. \\

As for the near-Earth spacecraft which could complete the network with a large, sensitive, all-sky GRB monitor, there are just two.  The Compton Gamma-Ray Observatory and INTEGRAL.  CGRO is now undergoing NASA's "Senior Review", which will determine its funding for the years 1999 - 2000, and make recommendations for the years 2001 - 2002.  These recommendations will be revisited in the year 2000.  Although there is no technical reason why the spacecraft could not operate beyond 2000, neither is there any guarantee of long-term support.  
\begin{figure}
\centerline{\epsfig{file=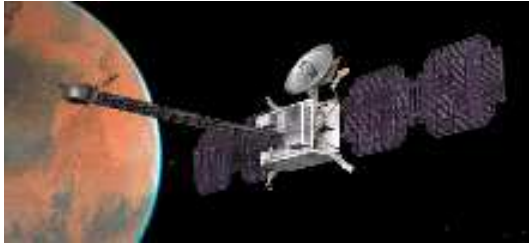, width=7cm}}
\caption{Figure 1. The Mars Surveyor Orbiter 2001 spacecraft.  The Ge spectrometer is mounted
on the boom; the neutron detector is on the spacecraft bus.
}
\end{figure}

\begin{figure}
\centerline{\epsfig{file=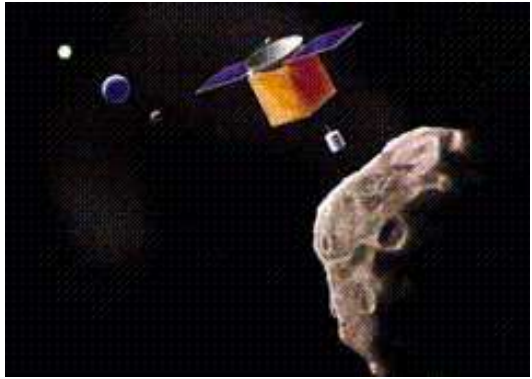, width=7cm}}
\caption{Figure 2. NEAP deploying a payload at Nereus.  The GRB experiment would remain on the
spacecraft bus.
}
\end{figure}

This leaves INTEGRAL.  At the urging of several people, the INTEGRAL SPI anticoincidence has been equipped with electronics which will allow it to be a very sensitive detector of gamma-ray bursts (Lichti 1998).  The anticoincidence shield will record and transmit count rates every 50 ms, time-tagged to an accuracy of 1 ms.  Lichti (1998) estimates the GRB detection rate to be 316 GRBs/y at the 5$\sigma$ level, and 160 GRBs/y at the 10$\sigma$ level.  

\bsk
\ni 3. CONCLUSIONS 
\ssk
\ni

INTEGRAL will have about the same burst detection rate as BATSE, and many of the characteristics of this future IPN can be estimated directly from the current one.  Assuming that we have at least two distant spacecraft in the network,

1) Approximately 70 GRBs/year can be detected and triangulated, increasing the total number of bursts from INTEGRAL by a factor of 3 - 4 (depending on whether the IPN bursts are the same as the IBIS ones),

2) If the network contains only two distant spacecraft, the resulting localizations will consist of alternate error boxes (the intersections of two triangulation annuli); however, in most cases, one can be rejected by considering the directional response of one or more of the detectors.  If it contains three distant spacecraft, each single GRB error box will be redundantly determined,

3) The error box sizes will range from 10's of arcseconds to degrees, with an average in the several arcminute range (e.g. Hurley et al. 1998),

4) The delays in deriving these error boxes will be determined by telemetry passes for the interplanetary spacecraft, and will range from hours in the very best cases, to a day or so in the worst cases.  However, recent work on two bursts, GRB980326 and 980703, has demonstrated that even delays of about one day are still adequate for identifying multiwavelength counterparts 

5) This network will bridge the gap until a dedicated MIDEX GRB mission can be launched in 2003, maintaining interest and expertise in GRB research.

\bsk
\baselineskip = 12pt
{\abstract \ni ACKNOWLEDGMENTS
The author is grateful for Ulysses support under JPL Contract 958056
}

\bsk
\baselineskip = 12pt

% beginning of font "references"

{\references \ni REFERENCES
\ssk

\ref Hurley, K., et al. 1998, The Ulysses Supplement to the BATSE 3B Catalog, ApJS, in press

\ref Lichti, G. 1998, MPE Document SPI-MPE-SP-1-19, Requirements Document for Gamma-Ray
Burst Measurements

\ref Pedersen, H. et al. 1997, in The Transparent Universe, ESA SP-382, p. 433

}                      

\end{document}